\documentclass{article}
\usepackage{amssymb}
\usepackage{amsmath}

\input{tcilatex}

\begin{document}

\title{Exact second-order structure-function relationships}
\author{Reginald J. Hill}
\maketitle

\begin{abstract}
Equations that follow from the Navier-Stokes equation and incompressibility
but with no other approximations are \textquotedblleft
exact.\textquotedblright . \ Exact equations relating second- and
third-order structure functions are studied, as is an exact
incompressibility condition on the second-order velocity structure function.
\ Opportunities for investigations using these equations are discussed. \
Precisely defined averaging operations are required to obtain exact averaged
equations. \ Ensemble, temporal, and spatial averages are all considered
because they produce different statistical equations and because they apply
to theoretical purposes, experiment, and numerical simulation of turbulence.
\ Particularly simple exact equations are obtained for the following cases:
i) the trace of the structure functions, ii) DNS that has periodic boundary
conditions, and iii)\ an average over a sphere in $\mathbf{r}$-space. \ The
last case (iii) introduces the average over orientations of $\mathbf{r}$\
into the structure function equations. \ The energy dissipation rate $%
\varepsilon $ appears in the exact trace equation without averaging, whereas
in previous formulations $\varepsilon $ appears after averaging and use of
local isotropy. \ The trace mitigates the effect of anisotropy in the
equations, thereby revealing that the trace of the third-order structure
function is expected to be superior for quantifying asymptotic scaling laws.
\ The orientation average has the same property.
\end{abstract}

\section{Introduction}

\qquad Equations relating statistics for turbulence studies, such as
Kolmogorov's (1941) equation are asymptotic equations. \ This has required
experimenters to seek turbulence that satisfies the criteria of the
asymptotic state. \ The present approach is to derive exact statistical
equations. \ These can be used to determine all effects contributing to the
balance of statistical equations. \ By \textquotedblleft
exact\textquotedblright\ we mean that the equations follow from the
Navier--Stokes equation and\ the incompressibility condition with no
additional approximations.

\qquad Exact equations have the potential to detect the limitations of
direct numerical simulation\ (DNS) and of experiments and to study the
approach to local homogeneity and local isotropy and scaling laws (Hill
2001). \ For those purposes, the averaging operation must be exactly defined
and implemented; that is done here. \ The methods developed here can be used
on the exact structure-function equations of all orders $N$; those equations
are in Hill (2001). \ It is useful to further investigate the exact
second-order ($N=2$) equation, which relates second- and third-order
structure functions, because it has special simplifications that the
higher-order equations ($N>2$) do not possess and because the second-order
equation is the most familiar. \ Exact equations satisfy the need perceived
by Yaglom (1998) for careful derivation of dynamic-theory equations and the
perceived value placed by Sreenivasan \& Antonia (1997) on aspects of
turbulence that can be understood precisely. \ Experimental data have been
used to evaluate the balance of Kolmogorov's equation (Antonia, Chambers \&
Browne 1983; Chambers \& Antonia 1984) and generalizations of it (Lindborg
1999; Danaila \textit{et al.} 1999 a,b; Antonia \textit{et al.} 2000). \
This report supports such experimental work as well as precise use of DNS by
giving exact equations to be used in such evaluations. \ The connection
between the derivations presented here and any experiment or DNS is
important because the equations relate several statistics and therefore are
most revealing when data are substituted into them.

\qquad The equations derived here are exact for every flow, whether laminar
or turbulent, provided that no forces act on the fluid at the points of
measurement, which points are denoted below by $\mathbf{x}$ and $\mathbf{x}%
^{\prime }$. \ Forces can be applied near the point of measurement; e.g.,
the equations are exact for hot-wire anemometer supports just downstream of
the measurement points. \ The cases of forces at the points of measurement
and throughout the fluid are considered in Hill (2002), wherein the
Kolmogorov flow computed by Borue \& Orszag (1996) is used as a specific
example.

\qquad The ensemble average is typically used for theoretical studies, the
temporal average for experimental data, and the spatial average for data
from DNS; thus all three are employed here. \ Ensemble, time and space
averages are not interchangeable because the averages commute differently
with differential operators within the dynamical equations. \ For the
homogeneous case and infinite averaging volume, the spatially averaged
equation (\ref{spatialaveext2}) and the ensemble averaged equation (\ref%
{exactave}) reduce to the same form, and similarly for the temporally
averaged equation for the stationary case and infinite averaging time.

\qquad Ongoing interest in turbulence intermittency includes accurate
evaluation of inertial-range exponents of structure functions, for which
purpose precise definition of an observed inertial range is needed. \ The
third-order structure function can serve this purpose because it has a
well-known inertial-range power law and the 4/5 coefficient (Kolmogorov
1941). \ Deviations from the 4/5 coefficient are observed in experiments
(Anselmet, Gagne \& Hopfinger 1984, observe values greater than 4/5 in duct
flow and less than 4/5 in jet flow despite the Reynolds numbers being nearly
equal; Mydlarski \& Warhaft 1996 and Lindborg 1999 obtain values less than
4/5, etc.); this casts doubt on the precision with which measured exponents
apply to the intermittency phenomenon (Sreenivasan \& Dhruva, 1998). \ The
equations derived here, when evaluated with data, can reveal all effects
contributing to the deviation from Kolmogorov's 4/5 law and his equation.

\qquad The plan of the paper is to develop the mathematics in \S 2 and \S 3; 
\S 2 contains necessary definitions and unaveraged equations; \S 3.1-3.3
contain the definition of averaging operations and their application to
produce averaged equations. \ \S 3.4 contains the simplifications for the
case of spatially periodic DNS. \ \S 3.5 defines the sphere average in $%
\mathbf{r}$\ space and its associated orientation average and relates these
to the work of Kolmogorov (1962) and Obukhov (1962); Kolmogorov's equation
is derived in \S 3.6 as a useful point of reference. \ Discussion of
opportunities that these equations present for future investigations is in 
\S 4.

\section{Exact unaveraged two-point equations}

\qquad The equations given here relate two-point quantities and are obtained
from the Navier--Stokes equations and incompressibility. \ The two spatial
points are denoted $\mathbf{x}$ and $\mathbf{x}^{\prime }$;\ they are
independent variables. \ They have no relative motion; e.g., anemometers at $%
\mathbf{x}$ and $\mathbf{x}^{\prime }$ are fixed relative to one another. \
To be concise, velocities are denoted $u_{i}=u_{i}(\mathbf{x},t)$, $%
u_{i}^{\prime }=u_{i}(\mathbf{x}^{\prime },t)$, energy dissipation rates by $%
\varepsilon =\varepsilon (\mathbf{x},t)$, $\varepsilon ^{\prime
}=\varepsilon (\mathbf{x}^{\prime },t)$ ,etc. \ $p$ is the pressure divided
by the density (density is constant), $\nu $ is kinematic viscosity, and $%
\partial $ denotes partial differentiation with respect to its subscript
variable. \ Summation is implied by repeated\ Roman indices; e.g., $\partial
_{x_{n}}\partial _{x_{n}}$ is the Laplacian operator. \ For brevity, define: 
\begin{eqnarray}
d_{ij} &\equiv &\left( u_{i}-u_{i}^{\prime }\right) \left(
u_{j}-u_{j}^{\prime }\right) , \\
d_{ijn} &\equiv &\left( u_{i}-u_{i}^{\prime }\right) \left(
u_{j}-u_{j}^{\prime }\right) \left( u_{n}-u_{n}^{\prime }\right) , \\
\tau _{ij} &\equiv &\left( \partial _{x_{i}}p-\partial _{x_{i}^{\prime
}}p^{\prime }\right) \left( u_{j}-u_{j}^{\prime }\right) +\left( \partial
_{x_{j}}p-\partial _{x_{j}^{\prime }}p^{\prime }\right) \left(
u_{i}-u_{i}^{\prime }\right) ,  \label{tau} \\
e_{ij} &\equiv &\left( \partial _{x_{n}}u_{i}\right) \left( \partial
_{x_{n}}u_{j}\right) +\left( \partial _{x_{n}^{\prime }}u_{i}^{\prime
}\right) \left( \partial _{x_{n}^{\prime }}u_{j}^{\prime }\right) ,
\label{eij} \\
\digamma _{ijn} &\equiv &\left( u_{i}-u_{i}^{\prime }\right) \left(
u_{j}-u_{j}^{\prime }\right) \frac{u_{n}+u_{n}^{\prime }}{2}.
\end{eqnarray}%
Change variables from $\mathbf{x}$ and $\mathbf{x}^{\prime }$ to the sum and
difference independent variables: 
\[
\mathbf{X}\equiv \left( \mathbf{x}+\mathbf{x}^{\prime }\right) /2\text{ \
and \ }\mathbf{r}\equiv \mathbf{x}-\mathbf{x}^{\prime }\text{, \ \ \ \ \ and
define }r\equiv \left\vert \mathbf{r}\right\vert .
\]%
The derivatives $\partial _{X_{i}}$ and $\partial _{r_{i}}$ are related to $%
\partial _{x_{i}}$ and $\ \partial _{x_{i}^{\prime }}$\ by$\ \ \ \ \ $%
\begin{equation}
\partial _{x_{i}}=\partial _{r_{i}}+\frac{1}{2}\partial _{X_{i}}\text{ , }\
\partial _{x_{i}^{\prime }}=-\partial _{r_{i}}+\frac{1}{2}\partial _{X_{i}}%
\text{ \ , }\partial _{X_{i}}=\partial _{x_{i}}+\partial _{x_{i}^{\prime }}%
\text{ \ , }\partial _{r_{i}}=\frac{1}{2}\left( \partial _{x_{i}}-\partial
_{x_{i}^{\prime }}\right) \text{.}  \label{derivs}
\end{equation}%
For any functions $f(\mathbf{x},t)$ and $g(\mathbf{x}^{\prime },t)$, (\ref%
{derivs}) gives 
\begin{equation}
\partial _{r_{i}}\left[ f(\mathbf{x},t)\pm g(\mathbf{x}^{\prime },t)\right]
=\partial _{X_{i}}\left[ f(\mathbf{x},t)\mp g(\mathbf{x}^{\prime },t)\right]
/2.  \label{identderivs}
\end{equation}

\qquad Use of (\ref{derivs}) in (\ref{tau}) and in the trace of (\ref{eij})
and rearranging terms gives

\begin{equation}
\tau _{ij}=-2\left( p-p^{\prime }\right) \left( s_{ij}-s_{ij}^{\prime
}\right) +\partial _{X_{i}}\left[ \left( p-p^{\prime }\right) \left(
u_{j}-u_{j}^{\prime }\right) \right] +\partial _{X_{j}}\left[ \left(
p-p^{\prime }\right) \left( u_{i}-u_{i}^{\prime }\right) \right] \text{,}
\label{tau2}
\end{equation}%
\begin{eqnarray}
e_{ii} &=&\nu ^{-1}\left( \varepsilon +\varepsilon ^{\prime }\right)
+\partial _{X_{n}}\partial _{X_{n}}\left( p+p^{\prime }\right) \text{,}
\label{eii} \\
\text{where \ \ \ \ \ \ \ \ \ \ \ \ \ \ }s_{ij} &\equiv &\left( \partial
_{x_{i}}u_{j}+\partial _{x_{j}}u_{i}\right) /2\text{ , and \ \ }\varepsilon
\equiv 2\nu s_{ij}s_{ij}\text{.\ \ \ \ \ \ \ \ \ \ \ \ \ \ \ \ \ \ }
\label{strain}
\end{eqnarray}%
To obtain (\ref{eii}) we used Poisson's equation, $\partial _{x_{n}}\partial
_{x_{n}}p=-\partial _{x_{i}}u_{j}\partial _{x_{j}}u_{i}$. \
Incompressibility requires $s_{ii}=0$; thus, the trace of (\ref{tau2}) is 
\begin{equation}
\tau _{ii}=2\partial _{X_{i}}\left[ \left( p-p^{\prime }\right) \left(
u_{i}-u_{i}^{\prime }\right) \right] .  \label{taucont}
\end{equation}%
Of course, all quantities above are local and instantaneous quantities.

\subsection{Use of the Navier--Stokes equation}

\qquad The Navier--Stokes equation and incompressibility give, 
\begin{equation}
\partial _{t}d_{ij}+\partial _{X_{n}}\digamma _{ijn}+\partial
_{r_{n}}d_{ijn}=-\tau _{ij}+2\nu \left( \partial _{r_{n}}\partial
_{r_{n}}d_{ij}+\frac{1}{4}\partial _{X_{n}}\partial
_{X_{n}}d_{ij}-e_{ij}\right) .  \label{exact2}
\end{equation}%
As a check, one sees that (\ref{exact2}) can be obtained from equation (2.9)
of Hill (2001). \ The trace of (\ref{exact2}) and substitution of (\ref{eii}%
) and (\ref{taucont}) gives 
\begin{equation}
\partial _{t}d_{ii}+\partial _{X_{n}}\digamma _{iin}+\partial
_{r_{n}}d_{iin}=2\nu \partial _{r_{n}}\partial _{r_{n}}d_{ii}-2\left(
\varepsilon +\varepsilon ^{\prime }\right) +w,  \label{exact2trace}
\end{equation}%
where 
\begin{equation}
w=-2\partial _{X_{i}}\left[ \left( p-p^{\prime }\right) \left(
u_{i}-u_{i}^{\prime }\right) \right] +\frac{\nu }{2}\partial
_{X_{n}}\partial _{X_{n}}d_{ii}-2\nu \partial _{X_{n}}\partial
_{X_{n}}\left( p+p^{\prime }\right) .  \label{ww}
\end{equation}%
The first term in (\ref{ww}) is $-\tau _{ii}$\ from (\ref{taucont}) and the
last term in (\ref{ww}) arises from $e_{ii}$\ in (\ref{eii}); the disparate
terms in (\ref{ww}) are given the symbol $w$\ for subsequent convenience and
brevity. \ The limit $r\rightarrow 0$ applied to (\ref{exact2trace})
recovers the definition of $\varepsilon $ in (\ref{strain}).

\subsection{Exact second-order incompressibility relationships}

\qquad Because $\mathbf{x}$ and $\mathbf{x}^{\prime }$\ are independent
variables,\ $\partial _{x_{i}}u_{j}^{\prime }=0$, and $\partial
_{x_{i}^{\prime }}u_{j}=0$. Then, incompressibility gives: $\partial
_{X_{n}}u_{n}=0,$ $\partial _{X_{n}}u_{n}^{\prime }=0$, $\partial
_{r_{n}}u_{n}=0$, $\partial _{r_{n}}u_{n}^{\prime }=0$, so $\partial
_{X_{n}}\left( u_{n}-u_{n}^{\prime }\right) =0$, and $\partial
_{r_{n}}\left( u_{n}-u_{n}^{\prime }\right) =0$. \ The combined use of those
incompressibility relations and (\ref{identderivs}) gives 
\begin{eqnarray}
\partial _{r_{n}}\left[ \left( u_{j}-u_{j}^{\prime }\right) \left(
u_{n}-u_{n}^{\prime }\right) \right] &=&\partial _{X_{n}}\left[ \left(
u_{j}+u_{j}^{\prime }\right) \left( u_{n}-u_{n}^{\prime }\right) \right] /2,
\label{secdorderincomp} \\
\partial _{r_{j}}\partial _{r_{n}}\left[ \left( u_{j}-u_{j}^{\prime }\right)
\left( u_{n}-u_{n}^{\prime }\right) \right] &=&\partial _{X_{j}}\partial
_{X_{n}}\left[ \left( u_{j}+u_{j}^{\prime }\right) \left(
u_{n}+u_{n}^{\prime }\right) \right] /4.  \label{secdorderincomp2}
\end{eqnarray}

\section{Exact averaged two-point equations}

\subsection{Ensemble average: \ exact equations}

\qquad The ensemble average is defined at each point $\left( \mathbf{X,r}%
,t\right) $ as the arithmetical average over the ensemble. \ We denote the
ensemble average by angle brackets and subscript $E$: $\left\langle \circ
\right\rangle _{E}$. \ Because the ensemble averaging operation is a
summation, it commutes with differential operators, the average of (\ref%
{exact2}) is 
\begin{eqnarray}
&&\partial _{t}\left\langle d_{ij}\right\rangle _{E}+\partial
_{X_{n}}\left\langle \digamma _{ijn}\right\rangle _{E}+\partial
_{r_{n}}\left\langle d_{ijn}\right\rangle _{E}  \nonumber \\
&=&-\left\langle \tau _{ij}\right\rangle _{E}+2\nu \left( \partial
_{r_{n}}\partial _{r_{n}}\left\langle d_{ij}\right\rangle _{E}+\frac{1}{4}%
\partial _{X_{n}}\partial _{X_{n}}\left\langle d_{ij}\right\rangle
_{E}-\left\langle e_{ij}\right\rangle _{E}\right) .  \label{exactave}
\end{eqnarray}
The argument list for each tensor in (\ref{exactave}) is $\left( \mathbf{X,r}%
,t\right) $; the ensemble average does not eliminate dependence on any
independent variable. \ The average of (\ref{exact2trace}) is 
\begin{equation}
\partial _{t}\left\langle d_{ii}\right\rangle _{E}+\partial
_{X_{n}}\left\langle \digamma _{iin}\right\rangle _{E}+\partial
_{r_{n}}\left\langle d_{iin}\right\rangle _{E}=2\nu \partial
_{r_{n}}\partial _{r_{n}}\left\langle d_{ii}\right\rangle _{E}-2\left\langle
\varepsilon +\varepsilon ^{\prime }\right\rangle _{E}+\left\langle
w\right\rangle _{E}.  \label{exactrace}
\end{equation}

\qquad Exact incompressibility conditions on the second-order velocity
structure function are given by the ensemble averages of (\ref%
{secdorderincomp}) and (\ref{secdorderincomp2}): 
\begin{eqnarray}
\partial _{r_{n}}\left\langle d_{jn}\right\rangle _{E} &=&\partial
_{X_{n}}\left\langle \left( u_{j}+u_{j}^{\prime }\right) \left(
u_{n}-u_{n}^{\prime }\right) \right\rangle _{E}/2,  \label{avesecdinc1} \\
\partial _{r_{j}}\partial _{r_{n}}\left\langle d_{jn}\right\rangle _{E}
&=&\partial _{X_{j}}\partial _{X_{n}}\left\langle \left( u_{j}+u_{j}^{\prime
}\right) \left( u_{n}+u_{n}^{\prime }\right) \right\rangle _{E}/4.
\label{avesecdinc2}
\end{eqnarray}

\subsection{Temporal average: \ exact equations}

\qquad Because nearly continuous temporal sampling is typical, we represent
the temporal average by an integral, but all results are valid for the sum
of discrete points as well. \ The temporal average is most useful when the
turbulence is nearly statistically stationary. \ Let $t_{0}$ be the start
time of the temporal average of duration $T$. \ The operator effecting the
temporal average of any quantity $Q$ is denoted by $\left\langle \circ
\right\rangle _{T}$, which has argument list $\left( \mathbf{X,r,}%
t_{0},T\right) $; that is, 
\begin{equation}
\left\langle Q\right\rangle _{T}\equiv \frac{1}{T}\int_{t_{0}}^{t_{0}+T}Q%
\left( \mathbf{X,r},t\right) dt.  \label{timeavedef}
\end{equation}%
The argument list $\left( \mathbf{X,r,}t_{0},T\right) $ is suppressed. \ The
temporal average of (\ref{exact2}--\ref{secdorderincomp2}) gives equations
that are the same form as (\ref{exactave}--\ref{avesecdinc2}) with one
exception. \ The exception is that $\partial _{t}$\ does not commute with
the integral operator (\ref{timeavedef}) such that $\left\langle \partial
_{t}d_{ij}\right\rangle _{T}$ appears, whereas $\partial _{t}\left\langle
d_{ij}\right\rangle _{E}$\ appears in (\ref{exactave}), and similarly for
the trace equation (\ref{exactrace}). \ Because data are taken at $\mathbf{x}
$ and $\mathbf{x}^{\prime }$ in the rest frame of the anemometers, and $%
\partial _{t}$ is the time derivative for that reference frame, it follows
that 
\begin{equation}
\left\langle \partial _{t}d_{ij}\right\rangle _{T}\equiv \frac{1}{T}%
\int_{t_{0}}^{t_{0}+T}\partial _{t}d_{ij}dt=\left[ d_{ij}\left( \mathbf{X,r}%
,t_{0}+T\right) -d_{ij}\left( \mathbf{X,r},t_{0}\right) \right] /T.
\label{tempavederiv}
\end{equation}%
This shows that it is easy to evaluate $\left\langle \partial
_{t}d_{ij}\right\rangle _{T}$\ using experimental data because only the
first (at $t=t_{0}$)\ and last (at $t=t_{0}+T$)\ data in the time series are
used. \ One can make $\left\langle \partial _{t}d_{ij}\right\rangle _{T}$ as
small as one desires by allowing $T$ to be very large provided that $%
d_{ij}\left( \mathbf{X,r},t_{0}+T\right) $ does not differ greatly from $%
d_{ij}\left( \mathbf{X,r},t_{0}\right) $. \ This is aided by judicious
choice of $t_{0}$\ and $t_{0}+T$\ for the stationary case, but is not
possible in all cases.

\subsection{Spatial average: \ exact equations}

\qquad Because nearly continuous spatial sampling is typical of DNS, we
represent the spatial average by an integral, but all results can be
generalized to the case of a sum of discrete points. \ Let the spatial
average be over a region $\mathbb{R}$ in $\mathbf{X}$-space. \ The spatial
average of any quantity $Q$ is denoted by $\left\langle Q\right\rangle _{%
\mathbb{R}}$ which has argument list $\left( \mathbf{r},t,\mathbb{R}\right) $%
; that is, 
\begin{equation}
\left\langle Q\right\rangle _{\mathbb{R}}\equiv \frac{1}{V}\underset{\mathbb{%
R}}{\int \int \int }Q\left( \mathbf{X,r},t\right) d\mathbf{X,}
\label{Xvolave}
\end{equation}
where $V$ is the volume of the space region $\mathbb{R}$. \ The argument
list $\left( \mathbf{r},t,\mathbb{R}\right) $ is suppressed. \ The spatial
average commutes with $\mathbf{r}$\ and $t$\ differential and integral
operations, and with ensemble, time, and $\mathbf{r}$-space\ averages, but
not with $\partial _{X_{n}}$. \ Given any vector $q_{n}$, the divergence
theorem relates the volume average of $\partial _{X_{n}}q_{n}$ to the
surface average; that is, 
\begin{equation}
\left\langle \partial _{X_{n}}q_{n}\right\rangle _{\mathbb{R}}\equiv \frac{1%
}{V}\int \int \int \partial _{X_{n}}q_{n}d\mathbf{X}=\frac{S}{V}\left( \frac{%
1}{S}\int \int \check{N}_{n}q_{n}dS\right) \equiv \frac{S}{V}\oint_{\mathbf{X%
}_{n}}q_{n},  \label{Xsurfave}
\end{equation}
where $S$ is the surface area bounding $\mathbb{R}$, $dS$ is the
differential of surface area, and $\check{N}_{n}$ is the unit vector
oriented outward and normal to the surface. \ As seen on the right-hand side
of (\ref{Xsurfave}), we adopt, for brevity, the integral-operator notation 
\[
\oint_{\mathbf{X}_{n}}\equiv \frac{1}{S}\int \int \check{N}_{n}dS.
\]

\qquad The spatial average of (\ref{exact2}) is 
\begin{eqnarray}
&&\partial _{t}\left\langle d_{ij}\right\rangle _{\mathbb{R}}+\frac{S}{V}%
\oint_{\mathbf{X}_{n}}\digamma _{ijn}+\partial _{r_{n}}\left\langle
d_{ijn}\right\rangle _{\mathbb{R}}  \nonumber \\
&=&-\left\langle \tau _{ij}\right\rangle _{\mathbb{R}}+2\nu \left( \partial
_{r_{n}}\partial _{r_{n}}\left\langle d_{ij}\right\rangle _{\mathbb{R}}+%
\frac{1}{4}\frac{S}{V}\oint_{\mathbf{X}_{n}}\partial
_{X_{n}}d_{ij}-\left\langle e_{ij}\right\rangle _{\mathbb{R}}\right) .
\label{spatialaveext2}
\end{eqnarray}
The spatial average of (\ref{exact2trace}) is 
\begin{equation}
\partial _{t}\left\langle d_{ii}\right\rangle _{\mathbb{R}}+\frac{S}{V}%
\oint_{\mathbf{X}_{n}}\digamma _{iin}+\partial _{r_{n}}\left\langle
d_{iin}\right\rangle _{\mathbb{R}}=2\nu \partial _{r_{n}}\partial
_{r_{n}}\left\langle d_{ii}\right\rangle _{\mathbb{R}}-2\left\langle
\varepsilon +\varepsilon ^{\prime }\right\rangle _{\mathbb{R}}+\left\langle
w\right\rangle _{\mathbb{R}},  \label{spatialavetrace}
\end{equation}
\[
\text{where\ \ \ }\left\langle w\right\rangle _{\mathbb{R}}\equiv \frac{S}{V}%
\oint_{\mathbf{X}_{n}}\left[ -2\left( p-p^{\prime }\right) \left(
u_{n}-u_{n}^{\prime }\right) +\frac{\nu }{2}\partial _{X_{n}}d_{ij}-2\nu
\partial _{X_{n}}\left( p+p^{\prime }\right) \right] .
\]

\qquad The spatial average of the incompressibility condition (\ref%
{secdorderincomp}) is 
\begin{equation}
\partial _{r_{n}}\left\langle d_{jn}\right\rangle _{\mathbb{R}}=\frac{S}{2V}%
\oint_{\mathbf{X}_{n}}\left( u_{n}-u_{n}^{\prime }\right) \left(
u_{j}+u_{j}^{\prime }\right) ,  \label{spacavescincop}
\end{equation}%
which is, on the right-hand side, a surface flux of a quantity that depends
on large-scale structures in the flow. \ Similarly, (\ref{secdorderincomp2})
gives 
\begin{equation}
\partial _{r_{j}}\partial _{r_{n}}\left\langle d_{jn}\right\rangle _{\mathbb{%
R}}=\frac{S}{4V}\oint_{\mathbf{X}_{n}}\partial _{X_{j}}\left[ \left(
u_{n}+u_{n}^{\prime }\right) \left( u_{j}+u_{j}^{\prime }\right) \right] .
\label{spaceavesincomp}
\end{equation}

\subsection{Spatial average: \ DNS with periodic boundary conditions}

\qquad The spatial average is particularly relevant to DNS. \ DNS that is
used to investigate turbulence at small scales often has periodic boundary
conditions. \ For such DNS, consider the spatial average over the entire DNS
domain.\ \ Contributions to $\oint_{\mathbf{X}_{n}}q_{n}$ from opposite
sides of the averaging volume cancel for that case such that $\oint_{\mathbf{%
X}_{n}}q_{n}=0$ and therefore $\left\langle \partial
_{X_{n}}q_{n}\right\rangle _{\mathbb{R}}=0$. \ In (\ref{spatialaveext2}) we
then have $\oint_{\mathbf{X}_{n}}\digamma _{ijn}=0$ and $\oint_{\mathbf{X}%
_{n}}\partial _{X_{n}}d_{ij}=0$. \ In (\ref{spatialavetrace}) we have $%
\oint_{\mathbf{X}_{n}}\digamma _{iin}=0$ and $\left\langle w\right\rangle _{%
\mathbb{R}}=0$. \ In (\ref{spacavescincop}), the right-hand side vanishes so
that $\partial _{r_{n}}\left\langle d_{jn}\right\rangle _{\mathbb{R}}=0$. \
Thus, in the DNS case described above, we have 
\begin{equation}
\partial _{t}\left\langle d_{ij}\right\rangle _{\mathbb{R}}+\partial
_{r_{n}}\left\langle d_{ijn}\right\rangle _{\mathbb{R}}=-\left\langle \tau
_{ij}\right\rangle _{\mathbb{R}}+2\nu \left( \partial _{r_{n}}\partial
_{r_{n}}\left\langle d_{ij}\right\rangle _{\mathbb{R}}-\left\langle
e_{ij}\right\rangle _{\mathbb{R}}\right) ,  \label{spaceDNS}
\end{equation}%
\begin{equation}
\partial _{t}\left\langle d_{ii}\right\rangle _{\mathbb{R}}+\partial
_{r_{n}}\left\langle d_{iin}\right\rangle _{\mathbb{R}}=2\nu \partial
_{r_{n}}\partial _{r_{n}}\left\langle d_{ii}\right\rangle _{\mathbb{R}%
}-2\left\langle \varepsilon +\varepsilon ^{\prime }\right\rangle _{\mathbb{R}%
},  \label{spaceDNStrace}
\end{equation}%
\begin{equation}
\partial _{r_{n}}\left\langle d_{jn}\right\rangle _{\mathbb{R}}=0\text{, and 
}\partial _{r_{n}}\left\langle e_{jn}\right\rangle _{\mathbb{R}}=0.
\label{spaceincomp2}
\end{equation}%
Proof of $\partial _{r_{n}}\left\langle e_{jn}\right\rangle _{\mathbb{R}}=0$
is given in Hill (2002).

\qquad Performing the $\mathbf{r}$-space divergence of (\ref{spaceDNS}) and
using (\ref{spaceincomp2}), we have 
\begin{equation}
\partial _{r_{j}}\partial _{r_{n}}\left\langle d_{ijn}\right\rangle _{%
\mathbb{R}}=-\partial _{r_{j}}\left\langle \tau _{ij}\right\rangle _{\mathbb{%
R}}.  \label{aaa}
\end{equation}
This exact result is analogous to the asymptotic result in Frisch (1995),
Lindborg (1996), and Hill (1997).

\qquad Using the Taylor series of $\varepsilon $\ and $\varepsilon ^{\prime
} $ around the point $\mathbf{X}$, Hill (2002) obtains the following exact
result for the periodic DNS case considered 
\begin{equation}
-2\left\langle \varepsilon +\varepsilon ^{\prime }\right\rangle _{\mathbb{R}%
}=-4\left\langle \varepsilon \left( \mathbf{X},t\right) \right\rangle _{%
\mathbb{R}},\text{ and }-\left\langle e_{ij}\right\rangle _{\mathbb{R}%
}=-4\nu \left\langle \left[ \left( \partial _{x_{n}}u_{i}\right) \left(
\partial _{x_{n}}u_{j}\right) \right] _{\mathbf{x}=\mathbf{X}}\right\rangle
_{\mathbb{R}},  \label{DNSeps}
\end{equation}%
where the subscript $\mathbf{x}=\mathbf{X}$ means that the derivatives are
evaluated at the point $\mathbf{X}$. \ \ An important aspect of (\ref{DNSeps}%
) is that the right-hand sides depend only on $t$. \ Of course, none of (\ref%
{spaceDNS}--\ref{aaa}) depends on $\mathbf{X}$ because of the spatial
average over $\mathbf{X}$.

\qquad No approximations have been used to obtain the above equations for
the spatially periodic DNS case considered.

\subsection{Averages over an $\mathbf{r}$-space sphere}

\qquad The energy dissipation rate averaged over a sphere in $\mathbf{r}$%
-space has been a recurrent theme in small-scale similarity theories since
its introduction by Obukhov (1962) and Kolmogorov (1962). \ By averaging our
equations for the trace, we can, for the first time, produce an exact
dynamical equation containing the sphere-averaged energy dissipation rate. \
The volume average over an $\mathbf{r}$-space sphere of radius $r_{S}$\ of a
quantity $Q$\ is denoted by 
\begin{equation}
\left\langle Q\right\rangle _{\mathbf{r}\text{-sphere}}\equiv \left( 4\pi
r_{S}^{3}/3\right) ^{-1}\underset{\left\vert \mathbf{r}\right\vert \text{ }%
\leq \text{ }r_{S}}{\int \int \int }Q\left( \mathbf{X,r},t\right) d\mathbf{r.%
}  \label{r-sphere}
\end{equation}%
The orientation average over the surface of the $\mathbf{r}$-space sphere of
radius $r_{S}$ of a vector $q_{n}\left( \mathbf{X,r},t\right) $ is denoted
by the following integral-operator notation: 
\begin{equation}
\oint_{\mathbf{r}_{n}}q_{n}\equiv \left( 4\pi r_{S}^{2}\right) ^{-1}\underset%
{\left\vert \mathbf{r}\right\vert \text{ }=\text{ }r_{S}}{\int \int }\frac{%
r_{n}}{r}q_{n}\left( \mathbf{X,r},t\right) ds,  \label{r-surface}
\end{equation}%
where $ds$ is the differential of surface area, and $r_{n}/r$ is the unit
vector oriented outward and normal to the surface of the $\mathbf{r}$-space
sphere. \ Note that $\left( 4\pi r_{S}^{2}\right) ^{-1}ds=d\Omega /4\pi $
where $d\Omega $\ is the differential of solid angle from the sphere's
center. \ Both $\left\langle Q\right\rangle _{\mathbf{r}\text{-sphere}}$\
and $\oint_{\mathbf{r}_{n}}q_{n}$\ have the argument list $\left( \mathbf{X,}%
r_{S},t\right) $, which is suppressed. \ The divergence theorem is 
\begin{equation}
\left\langle \partial _{r_{n}}q_{n}\right\rangle _{\mathbf{r}\text{-sphere}%
}=\left( 3/r_{S}\right) \oint_{\mathbf{r}_{n}}q_{n}\text{.}  \label{r-Gauss}
\end{equation}%
Because $\mathbf{r}$, $\mathbf{X}$, and $t$ are independent variables, the $%
\mathbf{r}$-space volume and orientation averages commute with time and $%
\mathbf{X}$-space averages and with $\mathbf{X}$- and $t$-differential
operators, and, of course, with the ensemble average as well. \ For
instance, 
\[
\left\langle \partial _{t}\left\langle d_{ii}\right\rangle _{\mathbb{R}%
}\right\rangle _{\mathbf{r}\text{-sphere}}=\partial _{t}\left\langle
\left\langle d_{ii}\right\rangle _{\mathbb{R}}\right\rangle _{\mathbf{r}%
\text{-sphere}}=\left\langle \left\langle \partial _{t}d_{ii}\right\rangle _{%
\mathbf{r}\text{-sphere}}\right\rangle _{\mathbb{R}}=\partial
_{t}\left\langle \left\langle d_{ii}\right\rangle _{\mathbf{r}\text{-sphere}%
}\right\rangle _{\mathbb{R}},\text{ etc.}
\]

\qquad The $\mathbf{r}$-sphere average (\ref{r-sphere}) can operate on all
of the above structure-function equations; it can operate on unaveraged
equations (\ref{exact2}) and (\ref{exact2trace}) as well. \ These equations
have terms of the form $\partial _{r_{n}}q_{n}$; e.g.,\ $q_{n}=\left\langle
d_{ijn}\right\rangle _{\mathbb{R}}$, $\partial _{r_{n}}\left\langle
d_{ii}\right\rangle _{\mathbb{R}}$, $\left\langle d_{iin}\right\rangle _{E}$%
, $\left\langle d_{ijn}\right\rangle _{T}$, $\partial _{r_{n}}\left\langle
d_{ii}\right\rangle _{T}$, etc. \ By means of (\ref{r-Gauss}), the volume
average in $\mathbf{r}$-space of any term of the form $\partial
_{r_{n}}q_{n} $ produces the orientation average of $q_{n}$ within the
subject equation. \ After operating on (\ref{exactrace}) with the volume
average in $\mathbf{r}$-space (\ref{r-sphere}), the term $-2\left\langle
\varepsilon +\varepsilon ^{\prime }\right\rangle _{E}$\ in that equation
produces $-2\left\langle \left\langle \varepsilon +\varepsilon ^{\prime
}\right\rangle _{\mathbf{r}\text{-sphere}}\right\rangle _{E}$. \ Now, $%
\left\langle \left\langle \varepsilon +\varepsilon ^{\prime }\right\rangle _{%
\mathbf{r}\text{-sphere}}\right\rangle _{E}/2$\ is the sphere-averaged
energy dissipation rate defined in the third equations of both Obukhov
(1962) and Kolmogorov (1962).

\qquad The result of the $\mathbf{r}$-space sphere average of any of our
equations will be clear from operating on (\ref{spaceDNStrace}). \ The
average of (\ref{spaceDNStrace}) over a sphere in $\mathbf{r}$-space of
radius $r_{S}$ and multiplication by $r_{S}/3$ and use of (\ref{DNSeps})
gives 
\begin{equation}
\frac{r_{S}}{3}\partial _{t}\left\langle \left\langle d_{ii}\right\rangle _{%
\mathbf{r}\text{-sphere}}\right\rangle _{\mathbb{R}}+\oint_{\mathbf{r}%
_{n}}\left\langle d_{iin}\right\rangle _{\mathbb{R}}=2\nu \oint_{\mathbf{r}%
_{n}}\partial _{r_{n}}\left\langle d_{ii}\right\rangle _{\mathbb{R}}-\frac{%
4r_{S}}{3}\left\langle \left\langle \varepsilon \right\rangle _{\mathbf{r}%
\text{-sphere}}\right\rangle _{\mathbb{R}}.  \label{simpleDNS}
\end{equation}%
The terms have argument list $\left( r_{S},t\right) $, but $\left\langle
\left\langle \varepsilon \right\rangle _{\mathbf{r}\text{-sphere}%
}\right\rangle _{\mathbb{R}}$\ depends only on $t$. \ Of course, none of the
quantities in (\ref{simpleDNS}) depends on $\mathbf{X}$ because of the $%
\mathbf{X}$-space average. \ Despite its simplicity, (\ref{simpleDNS}) has
been obtained without approximations for the freely-decaying
spatially-periodic DNS case considered; (\ref{simpleDNS})\ applies to
inhomogeneous and anisotropic DNS that have periodic boundary conditions. \
Nie \& Tanveer (1999) define a structure function $\widetilde{S}_{3}$ using
time, space, and solid-angle averages acting on $d_{iin}$, and consider the
asymptotic inertial range case to obtain that $\widetilde{S}_{3}=-\left(
4/3\right) \epsilon r$\ without use of local isotropy. \ An analogous result
can be obtained by applying inertial-range asymptotics to (\ref{simpleDNS});
namely, neglect the time-derivative term on the basis of local stationarity
and neglect the term proportional to $\nu $.

\subsection{Kolmogorov's equation derived from (\protect\ref{simpleDNS})}

\qquad Most readers are familiar with Kolmogorov's (1941) famous equation
that is valid for locally isotropic turbulence. \ A useful point of
reference is to derive it from (\ref{simpleDNS}). \ This helps elucidate (%
\ref{simpleDNS}). \ An index $1$ denotes projection in the direction of $%
\mathbf{r}$, and indices $2$ and $3$ denote orthogonal directions
perpendicular to $\mathbf{r}$. \ For locally isotropic turbulence we recall
that the only nonzero components of $\left\langle d_{ijn}\right\rangle _{%
\mathbb{R}}$ are $\left\langle d_{111}\right\rangle _{\mathbb{R}}$, $%
\left\langle d_{221}\right\rangle _{\mathbb{R}}=\left\langle
d_{331}\right\rangle _{\mathbb{R}}$, and of $\left\langle
d_{ij}\right\rangle _{\mathbb{R}}$ are$\ \left\langle d_{11}\right\rangle _{%
\mathbb{R}}$, and\ $\left\langle d_{22}\right\rangle _{\mathbb{R}%
}=\left\langle d_{33}\right\rangle _{\mathbb{R}}$. \ These components depend
only on $r$ such that there is no distinction in an $\mathbf{r}$-space
sphere average between $r_{S\text{ }}$ and $r$; thus, we simplify the
notation by replacing $r_{S\text{ }}$ with $r$. \ The isotopic-tensor
formula for $\left\langle d_{ijn}\right\rangle _{\mathbb{R}}$ gives $%
\left\langle d_{iin}\right\rangle _{\mathbb{R}}=\left( r_{n}/r\right) \left(
\left\langle d_{111}\right\rangle _{\mathbb{R}}+2\left\langle
d_{221}\right\rangle _{\mathbb{R}}\right) =\left( r_{n}/r\right)
\left\langle d_{ii1}\right\rangle _{\mathbb{R}}$, substitution of which into
(\ref{r-surface}) gives $\oint_{\mathbf{r}_{n}}\left\langle
d_{iin}\right\rangle _{\mathbb{R}}=\left( r_{n}/r\right) \left\langle
d_{iin}\right\rangle _{\mathbb{R}}=\left( r_{n}/r\right) \left(
r_{n}/r\right) \ \left\langle d_{ii1}\right\rangle _{\mathbb{R}}=$ \ $%
\left\langle d_{ii1}\right\rangle _{\mathbb{R}}$. Since $\left( \partial
_{r_{n}}r\right) =\left( r_{n}/r\right) $, we have $\oint_{\mathbf{r}%
_{n}}\partial _{r_{n}}\left\langle d_{ii}\right\rangle _{\mathbb{R}}=\left(
r_{n}/r\right) \left( \partial _{r_{n}}r\right) \partial _{r}\left\langle
d_{ii}\right\rangle _{\mathbb{R}}=\partial _{r}\left\langle
d_{ii}\right\rangle _{\mathbb{R}}$. \ Kolmogorov (1941) considered the
locally stationary case such that he neglected the time-derivative term,
thus we also neglect that term to obtain from (\ref{simpleDNS}) that 
\begin{equation}
\left\langle d_{ii1}\right\rangle _{\mathbb{R}}=2\nu \partial
_{r}\left\langle d_{ii}\right\rangle _{\mathbb{R}}-\frac{4}{3}\left\langle
\varepsilon \right\rangle _{\mathbb{R}}r.  \label{Kol1trace}
\end{equation}
Alternatively, we can time average (\ref{simpleDNS}); then the time
derivative can be neglected with the weaker condition noted with respect to
the smallness of (\ref{tempavederiv}); then $\left\langle \left\langle
d_{ii1}\right\rangle _{\mathbb{R}}\right\rangle _{T}=2\nu \partial
_{r}\left\langle \left\langle d_{ii}\right\rangle _{\mathbb{R}}\right\rangle
_{T}-\frac{4}{3}\left\langle \left\langle \varepsilon \right\rangle _{%
\mathbb{R}}\right\rangle _{T}r$. \ For simplicity of notation, we continue
with (\ref{Kol1trace}). \ To eliminate $\left\langle d_{22}\right\rangle _{%
\mathbb{R}}$ and $\left\langle \left\langle d_{221}\right\rangle _{\mathbb{R}%
}\right\rangle $ from the expressions $\left\langle d_{ii}\right\rangle _{%
\mathbb{R}}=\left\langle d_{11}\right\rangle _{\mathbb{R}}+2\left\langle
d_{22}\right\rangle _{\mathbb{R}}$ and $\left\langle d_{ii1}\right\rangle _{%
\mathbb{R}}=\left\langle d_{111}\right\rangle _{\mathbb{R}}+2\left\langle
d_{221}\right\rangle _{\mathbb{R}}$, we use the incompressibility conditions 
$\frac{r}{2}\partial _{r}\left\langle d_{11}\right\rangle _{\mathbb{R}%
}+\left\langle d_{11}\right\rangle _{\mathbb{R}}-\left\langle
d_{22}\right\rangle _{\mathbb{R}}=0$, and $r\partial _{r}\left\langle
d_{111}\right\rangle _{\mathbb{R}}+\left\langle d_{111}\right\rangle _{%
\mathbb{R}}-6\left\langle d_{221}\right\rangle _{\mathbb{R}}=0$, which are
valid for local isotropy (Hill 1997), and were also used by Kolmogorov
(1941). \ Then (\ref{Kol1trace}) becomes, after multiplying by $3r^{-1}$, $%
\partial _{r}\left\langle d_{111}\right\rangle _{\mathbb{R}}+\frac{4}{r}%
\left\langle d_{111}\right\rangle _{\mathbb{R}}=6\nu \left[ \partial
_{r}^{2}\left\langle d_{11}\right\rangle _{\mathbb{R}}+\frac{4}{r}%
\left\langle d_{11}\right\rangle _{\mathbb{R}}\right] -4\left\langle
\varepsilon \right\rangle _{\mathbb{R}}$; this is Kolmogorov's (1941) third
equation. \ After multiplication by $r^{4}$and integrating from $0$ to $r$,
we have Kolmogorov's (1941) equation 
\begin{equation}
\left\langle d_{111}\right\rangle _{\mathbb{R}}=6\nu \partial
_{r}\left\langle d_{11}\right\rangle _{\mathbb{R}}-\frac{4}{5}\left\langle
\varepsilon \right\rangle _{\mathbb{R}}r.  \label{koleq}
\end{equation}
Kolmogorov's inertial-range $4/5$ law and the viscous-range law follow
immediately from (\ref{koleq}).

\section{Examples of opportunities for using the exact equations}

\subsection{Mitigating anisotropy to check asymptotic laws\label{subhead}}

\qquad Consider homogeneous, anisotropic turbulence. \ Homogeneity causes $%
\partial _{X_{n}}$ operating on a statistic to vanish (Hill 2001), so $%
\partial _{X_{n}}\left\langle \digamma _{ijn}\right\rangle _{E}$ and $%
\partial _{X_{n}}\partial _{X_{n}}\left\langle d_{ij}\right\rangle _{E}$
vanish from (\ref{exactave}), but $\left\langle \tau _{ij}\right\rangle _{E}$
becomes $-2\left\langle \left( p-p^{\prime }\right) \left(
s_{ij}-s_{ij}^{\prime }\right) \right\rangle _{E}$ (see \ref{tau2}), which
does not vanish. \ Under the more restrictive assumption of local isotropy, $%
\left\langle \tau _{ij}\right\rangle _{E}=0$ (Hill 1997) such that the
entire nonzero value of $\left\langle \tau _{ij}\right\rangle _{E}$ is a
source of anisotropy in (\ref{exactave}). \ For the locally stationary case,
the anisotropy quantified by $\left\langle \tau _{ij}\right\rangle _{E}$ is
approximately balanced by that from the term $\partial _{r_{n}}\left\langle
d_{ijn}\right\rangle _{E}$ in (\ref{exactave}) (Hill 1997, and exactly so
for the stationary case). \ In contrast consider (\ref{exactrace}). \
Homogeneity causes $\partial _{X_{n}}\digamma _{iin}$ and $w$ to vanish from
(\ref{exactrace}); equivalently, $\left\langle \tau _{ii}\right\rangle _{E}$
is absent from (\ref{exactrace})\ because incompressibility gives $%
s_{ii}-s_{ii}^{\prime }=0$. \ Therefore, for the homogeneous, anisotropic
case, an important source of anisotropy of $\partial _{r_{n}}\left\langle
d_{ijn}\right\rangle _{E}$, namely $\left\langle \tau _{ij}\right\rangle
_{E} $, is absent from $\partial _{r_{n}}\left\langle d_{iin}\right\rangle
_{E}$. \ It therefore seems that $\left\langle d_{ii1}\right\rangle _{E}$
will more accurately show the asymptotic inertial-range power law than does $%
\left\langle d_{111}\right\rangle _{E}$ (or $\left\langle
d_{221}\right\rangle _{E}$\ or $\left\langle d_{331}\right\rangle _{E}$). \
This result for the homogeneous case extends to the locally homogeneous case
as follows: \ For inhomogeneous turbulence, the nonvanishing part of $%
\left\langle \tau _{ii}\right\rangle _{E}$, i.e., $\left\langle \tau
_{ii}\right\rangle _{E}=2\partial _{X_{i}}\left\langle \left( p-p^{\prime
}\right) \left( u_{i}-u_{i}^{\prime }\right) \right\rangle _{E}$ [see (\ref%
{taucont})] is expected to approach zero rapidly as $r$ decreases for two
reasons. \ First, $\left\langle \left( p-p^{\prime }\right) \left(
u_{i}-u_{i}^{\prime }\right) \right\rangle _{E}$ vanishes on the basis of
local isotropy. \ Second, the operator $\partial _{X_{i}}$ causes $\partial
_{X_{i}}\left\langle \left( p-p^{\prime }\right) \left( u_{i}-u_{i}^{\prime
}\right) \right\rangle _{E}$\ to vanish on the basis of local homogeneity. \
From (\ref{ww}), $\left\langle w\right\rangle _{E}$ contains the terms $\nu
\partial _{X_{n}}\partial _{X_{n}}\left\langle d_{ii}\right\rangle _{E}/2$
and $-2\nu \partial _{X_{n}}\partial _{X_{n}}\left\langle p+p^{\prime
}\right\rangle _{E}$; because of the operator $\partial _{X_{n}}\partial
_{X_{n}}$, these terms vanish on the basis of local homogeneity. \ Thus, all
terms in $\left\langle w\right\rangle _{E}$ are negligible for locally
homogeneous turbulence. \ By performing the trace, it appears that
anisotropy has been significantly reduced in (\ref{exactrace}) relative to
in (\ref{exactave}) for the high-Reynolds-number, locally homogeneous case
such that the above hypothesis is extended to locally homogeneous
turbulence. \ The hypothesis should be checked by comparison with
anisotropic DNS. \ Evaluation of all terms in (\ref{exactave}) and (\ref%
{exactrace}) are the basis for such an investigation. \ The above discussion
holds for temporal and spatial averages as well.

\qquad To determine scaling properties of the third-order structure
function, past theory has used the isotropic formulas. \ One can use an
equation like (\ref{exactrace}) or its temporal-average analogue without an
assumption about the symmetry properties (e.g., isotropic) of the structure
functions by means of the sphere average in $\mathbf{r}$-space. \ Without
approximation, the $\mathbf{r}$-space sphere average produces the
orientation-averaged third-order structure function. \ It would seem that
the orientation average mitigates anisotropy effects.\ \ Thus, the
orientation average $\oint_{\mathbf{r}_{n}}\left\langle d_{iin}\right\rangle
_{E}$ (or $\oint_{\mathbf{r}_{n}}\left\langle d_{iin}\right\rangle _{T}$, or 
$\oint_{\mathbf{r}_{n}}\left\langle d_{iin}\right\rangle _{\mathbb{R}}$) is
expected to best exhibit scaling properties of locally isotropic turbulence,
such as the inertial-range power law with the $4/3$ coefficient in (\ref%
{simpleDNS}). \ This hypothesis should be checked by comparison with
anisotropic DNS.

\qquad Consider the stationary, homogeneous case. \ From (\ref{exactrace}),
we are certain that $\partial _{r_{n}}\left\langle d_{iin}\right\rangle
_{E}-2\nu \partial _{r_{n}}\partial _{r_{n}}\left\langle d_{ii}\right\rangle
_{E}$ scales with $\left\langle \varepsilon +\varepsilon ^{\prime
}\right\rangle _{E}$ because 
\[
\left( \partial _{r_{n}}\left\langle d_{iin}\right\rangle _{E}-2\nu \partial
_{r_{n}}\partial _{r_{n}}\left\langle d_{ii}\right\rangle _{E}\right)
/\left\langle \varepsilon +\varepsilon ^{\prime }\right\rangle _{E}=-2;
\]%
thereby insuring K41 scaling of $\partial _{r_{n}}\left\langle
d_{iin}\right\rangle _{E}-2\nu \partial _{r_{n}}\partial
_{r_{n}}\left\langle d_{ii}\right\rangle _{E}$ despite anisotropy. \ In
contrast, (\ref{exactave}) insures that scaling only if local isotropy is
invoked. \ Anisotropic DNS can be used to check whether or not K41 scaling
is improved by performing the trace.

\subsection{Tests using DNS and experimental data}

\qquad The spatially periodic DNS case leads to especially simple equations.
\ It seems that (\ref{spaceDNS}--\ref{spaceDNStrace}) offer an ideal
opportunity to evaluate the contribution of $\partial _{t}\left\langle
d_{ij}\right\rangle _{\mathbb{R}}$ for freely decaying turbulence, and of $%
\left\langle \tau _{ij}\right\rangle _{\mathbb{R}}$ for anisotropic
turbulence, as well as the balance of the off-diagonal components of (\ref%
{spaceDNS}). \ Because we have not introduced a force generating the
turbulence and because every point in the flow enters into the $\mathbf{X}$%
-space average, the DNS must be freely decaying. \ As shown in Hill (2002),
it is straightforward to include forces in the equations. \ New experimental
methods (Su \& Dahm 1996) and DNS can completely evaluate terms in the exact
structure-function equations.

\subsection{Effect of inhomogeneity on incompressibility conditions}

\qquad Exact incompressibility relationships (\ref{avesecdinc1}--\ref%
{avesecdinc2}) are obtained that can be used to quantify the nonzero value
of $\partial _{r_{n}}\left\langle d_{jn}\right\rangle _{E}$\ (or of $%
\partial _{r_{n}}\left\langle d_{jn}\right\rangle _{T}$, or of $\partial
_{r_{n}}\left\langle d_{jn}\right\rangle _{\mathbb{R}}$) caused by
inhomogeneity. \ If inhomogeneity is only in the streamwise (say 1-axis)
direction, then the time average gives $\partial _{r_{n}}\left\langle
d_{jn}\right\rangle _{T}=\partial _{X_{1}}\left\langle \left(
u_{j}+u_{j}^{\prime }\right) \left( u_{1}-u_{1}^{\prime }\right)
\right\rangle _{T}/2$, which can be evaluated using anemometers. \ As $%
r\rightarrow 0$, (\ref{avesecdinc2}) becomes the second derivative with
respect to measurement location of the velocity variance and therefore
clearly depends on flow inhomogeneity.

\subsection{Quantifying effects of inhomogeneity and anisotropy on scaling
exponents}

\qquad Sreenivasan \& Dhruva (1998) note that one could determine scaling
exponents with greater confidence if one has a theory that exhibits not only
the asymptotic power law but also the trend toward the power law. \ Such a
theory must require difficult measurements or DNS to evaluate such trends.
The equations given here are the required theory for the third-order
structure function, given that data must be used to evaluate the equations
in a manner analogous to previously cited evaluations. \ In fact, it is not
possible that exact equations do not contain the physical effects discussed
by Sreenivasan \& Dhruva (1998). \ They discuss the fact that there is
correlation of velocity increments with large-scale velocity in
inhomogeneous turbulence, even for very large Reynolds numbers and $r$ in
the inertial range, but not so in isotropic turbulence. \ Our term $\partial
_{X_{n}}\left\langle \digamma _{iin}\right\rangle _{E}=\partial
_{X_{n}}\left\langle \left\vert \mathbf{u}-\mathbf{u}^{\prime }\right\vert
^{2}\left( u_{n}+u_{n}^{\prime }\right) /2\right\rangle _{E}$\ in (\ref%
{exactrace}), and all such analogous terms in the other equations,
explicitly contains such correlation, and the balance of the equations
imparts that correlation effect to the other statistics; all such terms do
vanish for isotropic turbulence. \ They also discuss the usefulness of
graphing all 3 terms in (\ref{koleq}) to discern the onset of the
dissipation-range. \ Our equations are exact there too.

\subsection{Quantifying effects of large-scale structure on small-scale
structure}

\qquad Experimenters remove the mean from an anemometer's signal before
calculating structure functions from the velocity fluctuations, whereas the
exact dynamical equations contain statistics of the full velocity field. \
Hill (2002) applied the Reynolds decomposition to the above exact dynamical
equations, and used inertial-range and viscous-range asymptotics to
determine the approximate dynamical equations pertaining to statistics of
fluctuations as well as all approximations that are required to obtain the
approximate equations. \ The Reynolds decomposition produces terms that
quantify the effect of the large-scale structure of turbulence on the small
scales. \ For example, $\partial _{X_{n}}\left\langle \digamma
_{ijn}\right\rangle _{E}$\ produces a generalization of the advective term
discovered by Lindborg (1999). \ Hill (2002) contrasts the various
definitions of local homogeneity and points out that the only definition
that simplifies dynamical equations is that from Hill (2001).

\ \ \ \ \ \ \ \ \ \ \ \ \ \ \ \ \ \ \ \ \ \ \ \ \ \ \ \ \ \ \ \ \ \ \ \
REFERENCES

\noindent \textsc{Anselmet, F., Gagne, E. J. \& Hopfinger, E. J.} 1984
High-order velocity structure functions in turbulent shear flows. \textit{J.
Fluid Mech}. \textbf{140}, 63--89.

\noindent \textsc{Antonia, R. A., Chambers, A. J. \& Browne, L. W. B.} 1983
Relations between structure functions of velocity and temperature in a
turbulent jet. \textit{Experiments in Fluids} \textbf{1}, 213--219.

\noindent \textsc{Antonia, R. A., Zhou, T., Danaila, L. \& Anselmet, F.}
2000 Streamwise inhomogeneity of decaying grid turbulence. \textit{Phys.
Fluids} \textbf{12}, 3086--3089.

\noindent \textsc{Borue, V. \& Orszag, S. A.} 1996 Numerical study of
three-dimensional Kolmogorov flow at high Reynolds numbers. \textit{J. Fluid
Mech}. \textbf{306}, 293--323.

\noindent \textsc{Chambers, A. J. \& Antonia, R. A.} 1984 Atmospheric
estimates of power-law exponents $\mu $\ and $\mu _{\theta }$. \textit{%
Bound.-Layer Meteorol.} \textbf{28}, 343--352.

\noindent \textsc{Danaila, L., Anselmet, F., Zhou, T. \& Antonia,\ R. A.}
1999a A generalization of Yaglom's equation which accounts for the
large-scale forcing in heated decaying turbulence.\textit{\ J. Fluid Mech.} 
\textbf{391}, 359--372.

\noindent \textsc{Danaila, L., Le Gal, P., Anselmet, F., Plaza, F. \&
Pinton, J. F.} 1999b Some new features of the passive scalar mixing in a
turbulent flow. \textit{Phys. Fluids} \textbf{11}, 636--646.

\noindent \textsc{Frisch, U.} 1995 \textit{Turbulence, The Legacy of A. N.
Kolmogorov}. Cambridge University Press.

\noindent \textsc{Hill, R. J.} 1997 Applicability of Kolmogorov's and
Monin's equations of turbulence. \textit{J.~Fluid Mech.} \textbf{353},
67--81.

\noindent \textsc{Hill, R. J.} 2001 Equations relating structure functions
of all orders. \textit{J. Fluid Mech. }\textbf{434}, 379--388.

\noindent \textsc{Hill, R. J.} 2002 The approach of turbulence to the
locally homogeneous asymptote as studied using exact structure-function
equations. (xxx.lanl.gov/physics/0206034).

\noindent \textsc{Kolmogorov, A. N.} 1941 Dissipation of energy in locally
isotropic turbulence. \textit{Dokl. Akad. Nauk SSSR} \textbf{32}, 16--18.

\noindent \textsc{Kolmogorov, A. N.} 1962\ A refinement of previous
hypotheses concerning the local structure of turbulence in a viscous
incompressible fluid at high Reynolds number. \textit{J.~Fluid Mech.} 
\textbf{13}, 82--85.

\noindent \textsc{Lindborg, E.} 1996 A note on Kolmogorov's third-order
structure-function law, the local isotropy hypothesis and the
pressure-velocity correlation. \textit{J.~Fluid Mech.} \textbf{326},
343--356.

\noindent \textsc{Lindborg, E.} 1999 Correction to the four-fifths law due
to variations of the dissipation. \textit{Phys. Fluids} \textbf{11},
510--512.

\noindent \textsc{Mydlarski, L. \& Warhaft, Z.} 1996 On the onset of
high-Reynolds-number grid-generated wind tunnel turbulence. \textit{J. Fluid
Mech.} \textbf{320}, 331--368.

\noindent \textsc{Nie, Q. \& Tanveer, S.} 1999 A note on third-order
structure functions in turbulence. \ \textit{Proc. Roy. Soc. Lond. A} 
\textbf{455}, 1615--1635.

\noindent \textsc{Obukhov, A. M.} 1962 Some specific features of atmospheric
turbulence. \textit{J. Fluid Mech. }\textbf{13}, 77--81.

\noindent \textsc{Su, L. K. \& Dahm, W. J. A.} 1996 Scalar imaging
velocimetry measurements of the velocity gradient tensor field in turbulent
flows. I. Experimental results. \textit{Phys. Fluids} \textbf{8}, 1883--1906.

\noindent \textsc{Sreenivasan, K. R. \& Antonia, R. A.} 1997\ The
phenomenology of small scale turbulence. \textit{Annu. Rev. Fluid Mech.} 
\textbf{29}, 435--472.

\noindent \textsc{Sreenivasan, K. R. \& Dhruva, B}. 1998 Is there scaling in
high-Reynolds-number turbulence? \textit{Prog. Theor. Phys. Suppl.} \textbf{%
130}, 103--120.

\noindent \textsc{Yaglom, A.} 1998 New remarks about old ideas of
Kolmogorov. \textit{Adv. Turb.} \textbf{VII}, 605--610.

\end{document}